\documentclass[preprint,showpacs,showkeys,preprintnumbers,amsmath,amssymb]{revtex4}
 \usepackage{dcolumn}
 \usepackage{bm}
 \usepackage{graphicx}

\begin{document}

 \title{Analytical study of superradiant instability of the five-dimensional
  Kerr-G\"{o}del black hole}

 \author{ Ran Li }

 \thanks{Electronic mail: liran.gm.1983@gmail.com}

 \affiliation{Department of Physics,
 Henan Normal University, Xinxiang 453007, China}

 \begin{abstract}

 We present the analytical study of superradiant instability
 of the rotating asymptotically G\"{o}del black hole
 (Kerr-G\"{o}del black hole) in five-dimensional minimal
 supergravity theory. By employing the matched asymptotic
 expansion method to solve Klein-Gordon equation of
 scalar field perturbation, we show that the complex parts
 of quasinormal frequencies are positive in the regime of
 superradiance. This implies the growing instability of
 superradiant modes. The reason for this kind of instability
 is the Dirichlet boundary condition at asymptotic infinity,
 which is similar to that of rotating black holes in
 anti-de Sitter (AdS) spacetime.

 \end{abstract}



 \maketitle

 \section{Introduction}

 Superradiance is a classical phenomenon associated with
 ergosphere in a rotating black hole
 \cite{zeldovich,bardeen,misner,starobinsky}. If the modes of
 impinging bosonic fields $\Phi\sim \exp\{-i\omega t+im\phi\}$,
 with frequency $\omega$ and angular momentum $m$, are scattered
 off by the event horizon of black hole, the requirement of
 superradiant amplification is $0<\omega<m\Omega_H$,
 where $\Omega_H$ is the angular velocity of event horizon.

 The superradiance phenomenon allows extracting the rotational
 energy efficiently from black hole. It has been proposed
 by Press and Teukolsky \cite{press} to use superradiance phenomenon
 to built the \emph{black-hole bomb}.
 The essential of black-hole bomb mechanism is to add a reflecting
 mirror outside the rotating black hole. Then superradiant modes
 will bounce back and forth between the event horizon and the mirror.
 Meanwhile, the rotational energy extracted from black hole by means of
 superradiance process will grow exponentially. This mechanism has
 been recently restudied by many authors \cite{cardoso2004bomb,Hod,Rosa,Lee}.

 When the reflecting mirror is not artificial, superradiance amplification
 of impinging wave can lead to the instability of black hole, which is just called
 superradiant instability. This kind of instability has been studied extensively
 in recent years. For example, Kerr and
 Kerr-Newman black holes \cite{kerrunstable,detweiler,dolan,hodPLB2012}
 and Kerr-Newman black hole immersed in magnetic field
 \cite{konoplyaPLB} are all unstable against the massive scalar field perturbations,
 where the mass terms of perturbations play the role of reflective mirrors.
 Five-dimensional boosted Kerr black string \cite{DiasPRD2006}
 is also unstable against the massless scalar field where Kaluza-Klein
 momentum works as reflective mirror.

 For the rotating black holes in AdS space,
 the boundary at infinity can also work as the reflecting mirror.
 Small Kerr-AdS black hole in four dimensions is unstable
 against massless scalar field \cite{cardoso2004ads} and
 gravitational field \cite{cardoso2006prd} perturbations.
 Contrary to four-dimensional case, superradiance instability of
 five-dimensional rotating charged AdS black hole \cite{aliev}
 occurs only when the orbital quantum number is even.
 More recently, superradiant instability of small Reissner-Nordstr\"{o}m-anti-de Sitter black hole
 is investigated analytically and numerically \cite{uchikata}.
 We also studied the superradiant instability of charged massive
 scalar field in Kerr-Newman-anti-de Sitter black hole \cite{ranliplb}.
 In fact, besides AdS space, there are also other cases
 where the boundary at asymptotic infinity provides the reflecting mirror.
 For example, the rotating linear dilaton black hole \cite{clement}
 and the charged Myers-Perry black hole in G\"{o}del
 universe \cite{knopolya} are unstable due to superradiance.
 The reason of superradiant instability for these black holes
 originates from the Dirichlet boundary condition at asymptotic infinity
 of the perturbation fields.

 As mentioned above, superradiant instability of charged
 rotating asymptotically G\"{o}del black hole has been found
 by using the numerical methods \cite{knopolya}. In this paper,
 we will \emph{re-investigate} the same aspect of this kind
 of rotating black holes in G\"{o}del universe by using the
 analytical methods. We focus on the rotating asymptotically
 G\"{o}del black hole in five-dimensional minimal
 supergravity theory. This black hole is also called as
 Kerr-G\"{o}del black hole in literature. Firstly, by considering
 the scalar field perturbation in this background,
 we find that the asymptotically G\"{o}del spacetime requires
 the wave equation to satisfy the Dirichlet boundary condition
 at asymptotic infinity. Then, we divide the space outside
 the event horizon of Kerr-G\"{o}del black hole into the
 near-region and the far-region, and employ the matched asymptotic
 expansion method to solve the wave equation of scalar field perturbation.
 We only deal with black hole in the limit of small rotating parameter $j$ of
 G\"{o}del universe. The analysis of complex quasinormal modes
 by imposing the boundary conditions shows that the complex parts
 are positive in the regime of superradiance,
 which implies the growing instability of
 these modes. This is to say that the five-dimensional Kerr-G\"{o}del black hole
 is unstable against the scalar field perturbation.
 The reason for this instability is just the Dirichlet boundary
 condition at asymptotic infinity, which is similar to that of the rotating black holes in
 AdS space.

 The remaining of this paper is arranged as follows. In Section 2,
 we give a brief review of Kerr-G\"{o}del black hole in five-dimensional
 minimal supergravity theory. In Section 3, we investigate the classical
 superradiance phenomenon and the boundary condition of scalar field
 perturbation. In Section 4, the approximated solution of wave equation
 for scalar field is obtained by using the matched asymptotic
 expansion method and the superradiant instability is explicitly shown.
 The last section is devoted to conclusion and discussion.

 \section{Five-dimensional Kerr-G\"{o}del black hole}

 The bosonic part of five-dimensional minimal supergravity theory
 consists of the metric and a one-form gauge field, which are governed
 by Einstein-Maxwell-Chern-Simons (EMCS) equations of motion
 \begin{eqnarray}
 &&R_{\mu\nu}-\frac{1}{2}R g_{\mu\nu}=2\left(
 F_{\mu\alpha}F_{\nu}^{\;\;\alpha}-\frac{1}{4}
 g_{\mu\nu}F_{\alpha\beta}F^{\alpha\beta}\right)
 \;,\nonumber\\
 &&D_\nu\left(F^{\mu\nu}+\frac{1}{\sqrt{3}\sqrt{-g}}\epsilon^{\mu\nu\alpha\beta\gamma}
 A_\alpha F_{\beta\gamma}\right)=0\;.
  \end{eqnarray}
 The five-dimensional Kerr-G\"{o}del black hole is a solution to the EMCS equations of motion,
 the metric of which takes the form \cite{gimon}
 \begin{eqnarray}\label{metric}
  ds^2&=&-f(r)dt^2-q(r)r\sigma_L^3dt-h(r)r^2(\sigma_L^3)^2+
  \frac{dr^2}{v(r)}\nonumber\\
  &&+\frac{r^2}{4}(d\theta^2+d\psi^2+d\phi^2+2\cos\theta
  d\psi d\phi)\;,
 \end{eqnarray}
 where $\sigma_L^3=d\phi+\cos\theta d\psi$, and the metric functions are given by
 \begin{eqnarray}
 f(r)&=&1-\frac{2M}{r^2}\;,\nonumber\\
 q(r)&=&2jr+\frac{2Ma}{r^3}\;,\nonumber\\
 h(r)&=&j^2(r^2+2M)-\frac{Ma^2}{2r^4}\;,\nonumber\\
 v(r)&=&1-\frac{2M}{r^2}+\frac{8jM(a+2jM)}{r^2}+\frac{2Ma^2}{r^4}\;.
 \end{eqnarray}

  The parameters $M$ and $a$ are related to the mass,
  and the angular momentum of black hole.
  In this metric, the parameter $j$ defines the scale of the G\"{o}del background
  and is responsible for the rotation of the G\"{o}del universe \cite{godel}.
  When $a=0$, this solution reduced to the Gimon-Hashimoto solution,
  i.e. the Schwarzschild black hole in G\"{o}del universe \cite{gimon}.
  The thermodynamics of this black hole has been studied in \cite{banichconpere,wupengprd}.
  The scalar field perturbation and greybody factor of Hawking radiation
  of this kind of black holes are also calculated in the limit of small $j$
  in \cite{scalargodel}. This black hole has also been generalized to
  being charged \cite{wu} and other forms.

  In this paper, we consider the non-extremal black hole case.
  The metric function $v(r)$ has two positive real roots $r_\pm$,
  which are given by
  \begin{eqnarray}\label{horizon}
  r_{\pm}^2=M-4jMa-8j^2M^2\pm\sqrt{\xi}\;,
  \end{eqnarray}
  where
  \begin{eqnarray}
  \xi=(M-4jMa-8j^2M^2)^2-2Ma^2\;.
  \end{eqnarray}
  Clearly the non-extremal condition is given by $\xi>0$.
  The event horizon locates at the largest root $r_+$ of function $v(r)$.

  For latter convenience, we also present the expression of angular momentum
  at the event horizon
  \begin{eqnarray}
  \Omega_H&=&\frac{2q(r_+)}{r_+(1-4h(r_+))}\nonumber\\
  &=&\frac{4(Ma+jr_+^4)}{r_+^4-4j^2r_+^4(r_+^2+2M)+2Ma^2}\;.
  \end{eqnarray}
  By employing the relation of metric functions
  \begin{eqnarray}
  q^2(r)+f(r)(1-4h(r))=v(r)\;,
  \end{eqnarray}
  and noting that $v(r)$ vanishes at the horizon,
  another simple expression for the angular momentum can be derived
  \begin{eqnarray}
  \Omega_H=\frac{2M-r_+^2}{Ma+jr_+^4}\;.
  \end{eqnarray}
  In the present paper, we consider the rotating parameters $a$ and $j$
  are both positive. From the expression of $r_+^2$ in (\ref{horizon}), one can easily find
  $r_+^2<2M$, which implies that the angular momentum $\Omega_H$ is always
  positive when the rotating parameters $a$ and $j$ are positive.

  \section{Superradiance and boundary condition}

 Now let us consider the wave equation of massless scalar field perturbation
 in the background (\ref{metric}), which is given by Klein-Gordon equation
 \begin{eqnarray}\label{waveequation}
  \nabla_\mu\nabla^\mu\Phi=\frac{1}{\sqrt{-g}}
  \partial_\mu(g^{\mu\nu}\sqrt{-g}\partial_\nu\Phi)=0\;.
 \end{eqnarray}

 Because the metric has the killing vectors $\partial_t$, $\partial_\psi$,
 and $\partial_\phi$, we can take the ansatz of scalar field as
 \begin{eqnarray}
  \Phi=e^{-i\omega t+in\psi+im\phi}\Theta(\theta)R(r)\;.
 \end{eqnarray}
 Substituting this ansatz into the wave equation (\ref{waveequation})
 and separating the variables, we can get the angular equation
 \begin{eqnarray}\label{angular}
 \frac{1}{\sin\theta}\partial_\theta\left[\sin\theta\partial_\theta\Theta(\theta)\right]
 -\frac{(n-m\cos\theta)^2}{\sin^2\theta}\Theta(\theta)
 +[l(l+1)-m^2]\Theta(\theta)=0\;,
 \end{eqnarray}
 and the radial equation
 \begin{eqnarray}\label{radial}
 \frac{1}{4r}\partial_r\left[r^3v(r)\partial_rR(r)\right]
 +\frac{r^2(1-4h(r))}{4v(r)}
 \left[\omega-\frac{2mq(r)}{r(1-4h(r))}\right]^2R(r)
 \nonumber\\
 +\left[l(l+1)+\frac{4m^2h(r)}{(1-4h(r))}\right]R(r)=0\;.
 \end{eqnarray}

 Obviously, the angular equation (\ref{angular}) is independent of the black hole parameters
 and is exactly solvable. The solutions for the angular equation are just
 the spin-weighted spherical harmonics functions, where the integers $l = 0, 1, 2, \cdots$ are
 the separation constants and the modes $m = 0,\pm 1, \cdots ,\pm l$.

 In the following, we will specify the appropriate boundary
 conditions for the instability problem.
 At the horizon, the third set of terms in radial wave equation (\ref{radial})
 can be neglected and this equation can be reduced to the form
 \begin{eqnarray}\label{radial1}
 v\partial_r(v\partial_rR(r))+(1-4h(r_+))(\omega-m\Omega_H)^2R(r)=0\;.
 \end{eqnarray}
 Near the horizon, we use the approximation $v(r)\cong 2(r_+^2-r_-^2)(r-r_+)/r_+^3$.
 Then the solution of equation (\ref{radial1}) satisfying the ingoing boundary condition
 at the horizon is given by
 \begin{eqnarray}
 R(r)\sim (r-r_+)^{-i\varpi}=e^{-i\varpi\ln(r-r_+)}\;,
 \end{eqnarray}
 where we have defined
 \begin{eqnarray}\label{varpi}
 \varpi=\frac{r_+(r_+^4+2Ma^2-4j^2r_+^6-8j^2Mr_+^2)^{1/2}}
 {2(r_+^2-r_-^2)}(\omega-m\Omega_H)\;.
 \end{eqnarray}

 This solution gives us the superradiance condition of five-dimensional
 Kerr-G\"{o}del black hole. When the frequency of the wave
 is such that $\varpi$ is negative, i.e. $\omega<m\Omega_H$,
 one is in the superradiant regime, and the amplitude of
 an ingoing bosonic field is amplified after scattering
 by the event horizon. Meanwhile, for the present purpose, it is enough to
 consider the frequency $\omega$ is positive, which gives the
 superradiance condition as
 \begin{eqnarray}
 0<\omega<m\Omega_H\;.
 \end{eqnarray}
 From this condition, one can see that superradiance will
 occur only for the positive $m$. In the following, we will
 only work with the positive $m$.

 At infinity, the radial wave equation (\ref{radial}) is dominated by
 \begin{eqnarray}
 r\partial_r^2 R(r)+3\partial_r R(r)-4j^2\omega^2 r^3 R(r)=0\;.
 \end{eqnarray}
 The solution is given by
 \begin{eqnarray}
 R(r)\sim \frac{1}{r^2}e^{-j\omega r^2}\;,
 \end{eqnarray}
 where we have used the analogy with AdS backgrounds and imposed
 the Dirichlet boundary conditions at spatial infinity.

 With the boundary conditions that the ingoing wave at the horizon and the Dirichlet boundary condition at the infinity, one can solve the complex quasinormal modes of the massless scalar field
 in Kerr-G\"{o}del background. If the imaginary part of quasinormal mode is negative, it is known that
 the system is stable against this kind of perturbation. The instability means that the imaginary
 part is positive. In the next section, we will calculate the quasinormal modes by using the
 matching technique. It is shown that, in the regime of superradiance, the imaginary part
 of quasinormal mode is positive. In other words, the superradiant instability
 of five-dimensional Kerr-G\"{o}del black hole can be found analytically.

 \section{Analytical calculation of superradiant instability}

 In this section, we will present an analytical calculation of
 superradiant instability for the massless scalar perturbation.
 We will adopt the so-called matched asymptotic expansion method
 to solve the radial wave equation (\ref{radial}). It turns out to be
 convenient to use the new variable $x$ defined by
 $x=r^2$. Then the radial wave equation can be transformed into
 \begin{eqnarray}\label{radialx}
 \Delta\partial_x(\Delta\partial_x)R(x)+\frac{x^3}{4}(1-4h(x))\left[
 \omega-\frac{2mq(x)}{\sqrt{x}(1-4h(x))}\right]^2R(x)
 \nonumber\\
 +\Delta\left[l(l+1)+\frac{4m^2h(x)}{1-4h(x)}\right]R(x)=0\;,
 \end{eqnarray}
 where we have used $\Delta=x^2v(x)=(x-x_+)(x-x_-)$ with $x_{\pm}=r_{\pm}^2$.

 In order to employ the matched asymptotic expansion method, we should
 take the assumption $\omega M\ll 1$, and divide
 the space outside the event horizon into two regions, namely, a near-region,
 $x-x_+\ll 1/\omega$, and a far-region, $x-x_+\gg M$. The approximated solution
 can be obtained by matching the near-region solution and the far-region solution
 in the overlapping region $M\ll x-x_+\ll1/\omega$.

 Previous numerical works \cite{konoplyaads,knopolya} on the spectrum of asymptotically G\"{o}del
 black holes show a number of common features with the spectrum of AdS spacetime, where
 the rotational parameter $j$ of G\"{o}del universe plays the role of the inverse AdS radius
 $\ell$. Inspired by the work of \cite{cardoso2004ads}, where small AdS black hole are considered,
 we will deal with the rotating asymptotically G\"{o}del black hole in the limit
 of small rotational parameter $j$ in the following. The small AdS black hole condition
 implies that $r_+/\ell\ll 1$. For the small G\"{o}del black hole,
 we assume that $jr_+\ll 1$.

 With these assumptions, we can analyse the properties of the solution and study the stability of black hole against the perturbation by imposing the appropriate boundary conditions obtained in the last section.

 \subsection{Near-region solution}

 Firstly, Let us focus on the near-region in the vicinity of the event horizon,
 $\omega(x-x_+)\ll 1$. For the small $j$ black holes, this means $jr_+\ll 1$.
 The radial wave function (\ref{radialx}) in the near-region can be reduced to the form
 \begin{eqnarray}
 \Delta\partial_x(\Delta\partial_xR(x))
 +\left[(x_+-x_-)^2\varpi^2
 -l(l+1)\Delta\right]R(r)=0\;.
 \end{eqnarray}
 Noted that the last term in Eq.(\ref{radialx}) is neglected because we
 only consider the case $m\sim\omega$.

 Introducing the new coordinate variable
 \begin{eqnarray}
 z=\frac{x-x_+}{x-x_-}\;,
 \end{eqnarray}
 the near-region radial equation can be written in the form of
 \begin{eqnarray}
 z\partial_z(z\partial_z R(z))
 +\left[\varpi^2-l(l+1)\frac{z}{(1-z)^2}\right]R(z)=0\;,
 \end{eqnarray}
 with
  \begin{eqnarray}
 \varpi=\frac{r_+(r_+^4+2Ma^2)^{1/2}}{2(r_+^2-r_-^2)}(\omega-m\Omega_H)\;.
 \end{eqnarray}
 This expression for $\varpi$ is coincide with the expression given in
 (\ref{varpi}) in the small $j$ limit.

 Through defining
 \begin{eqnarray}
 R=z^{i\varpi}(1-z)^{l+1}F(z)\;,
 \end{eqnarray}
 the near-region radial wave equation becomes
 \begin{eqnarray}
 z(1-z)\partial_z^2F(z)+[c-(1+a+b)]\partial_zF(z)-abF(z)=0\;,
 \end{eqnarray}
 with the parameters
 \begin{eqnarray}
 a&=&l+1+2i\varpi\;,\nonumber\\
 b&=&l+1\;,\nonumber\\
 c&=&1+2i\varpi\;.
 \end{eqnarray}

 In the neighborhood of $z=0$, the general solution of
 the radial wave equation is given in terms of the hypergeometric function
 \begin{eqnarray}
  R&=&Az^{-i\varpi}(1-z)^{l+1}F(l+1,l+1-2i\varpi,1-2i\varpi,z)
  \nonumber\\
  &&+Bz^{i\varpi}(1-z)^{l+1}F(l+1,l+1+2i\varpi,1+2i\varpi,z)\;.
 \end{eqnarray}
 It is obvious that the first term represents the ingoing wave
 at the horizon, while the second term represents the outgoing
 wave at the horizon. Because we are considering the classical
 superradiance process, the ingoing boundary condition at the
 horizon should be employed. Then we have to set $B=0$. The physical solution of
 the radial wave equation corresponding to the ingoing wave
 at the horizon is then given by
 \begin{eqnarray}
 R=Az^{-i\varpi}(1-z)^{l+1}F(l+1,l+1-2i\varpi,1-2i\varpi,z)\;.
 \end{eqnarray}

 In order to match the far-region solution that will be obtained
 in the next subsection, we should study the large $r$, $z\rightarrow 1$,
 behavior of the near-region solution. For the sake of this purpose, we can
 use the $z\rightarrow 1-z$ transformation law for the hypergeometric function
 \begin{eqnarray}
 F(a,b,c,z)&=&\frac{\Gamma(c)\Gamma(c-a-b)}{\Gamma(c-a)\Gamma(c-b)}
 F(a,b,a+b-c+1,1-z)\nonumber\\
 &&+(1-z)^{c-a-b}
 \frac{\Gamma(c)\Gamma(a+b-c)}{\Gamma(a)\Gamma(b)}\nonumber\\
 &&\times
 F(c-a,c-b,c-a-b+1,1-z)\;\;.
 \end{eqnarray}
 By employing this formula and using the properties of hypergeometric function
 $F(a,b,c,0)=1$, we can get the large $r$ behavior of the near-region solution as
 \begin{eqnarray}\label{nearsolutionlarge}
 R&\sim& A\Gamma(1-2i\varpi)\left[\frac{(r_+^2-r_-^2)^{-l}\Gamma(2l+1)}
 {\Gamma(l+1)\Gamma(l+1-2i\varpi)}r^{2l}\right.
 \nonumber\\&&\left.
 +\frac{(r_+^2-r_-^2)^{l+1}\Gamma(-2l-1)}
 {\Gamma(-l)\Gamma(-l-2i\varpi)}r^{-2l-2}\right]\;,
 \end{eqnarray}
 where the variable $x$ has been restored to $r$ for later convenience.
 This solution should be matched with the small $r$ behavior of the far-region solution
 obtained in the next subsection.

 \subsection{Far-region solution}

 In the Far-region, $x-x_+\gg M$, we can neglect the effects induced by the black hole,
 i.e. we have $a\sim 0$ and $M\sim 0$. The metric functions can be approximated as
 $v(x)=f(x)=1$, $h(x)=j^2x$, and $q(x)=2j\sqrt{x}$.
 One can deduce the far-region radial wave equation as
 \begin{eqnarray}
 \partial_x^2(xR)+\left[-j^2\omega^2+\frac{\omega(\omega-8mj)}{4x}-\frac{l(l+1)}{x^2}\right]
 (xR)=0\;.
 \end{eqnarray}
 By defining the new variable $\zeta=2j\omega x$, the
 far-region radial wave equation can be reduced to
 \begin{eqnarray}
 \partial_\zeta^2(\zeta R)+\left[-\frac{1}{4}+\frac{\rho}{\zeta}
 -\frac{l(l+1)}{\zeta^2}\right](\zeta R)=0\;,
 \end{eqnarray}
 with the parameter $\rho=(\omega-8mj)/8j$.

 This is a standard Whittaker equation $\partial_\zeta^2 W+
 [-1/4+\rho/\zeta+(1/4-\mu^2)/\zeta^2]W=0$ with $W=\zeta R$ and
 $\mu=l+1/2$. The general solution is given by $W=\zeta^{\mu+1/2}
 e^{-\zeta/2}[\alpha M(\tilde{a},\tilde{b},\zeta)+\beta
 U(\tilde{a},\tilde{b},\zeta)]$, where $M$ and $U$ are Whittaker's
 functions with $\tilde{a}=1/2+\mu-\rho$ and $\tilde{b}=1+2\mu$.
 So the far-region solution of the radial wave equation is given by
 \begin{eqnarray}
 R=\zeta^l e^{-\zeta/2}\left[\alpha M(l+1-\rho,2l+2,\zeta)
 +\beta U(l+1-\rho,2l+2,\zeta)\right]\;.
 \end{eqnarray}

 Now we want to impose the boundary condition at asymptotic infinity.
 We are interested in the superradiance region with $0<\omega<m\Omega_H$,
 so we have $\zeta=2j\omega r^2\rightarrow +\infty$ when $r\rightarrow+\infty$.
 When $\zeta\rightarrow+\infty$, by using the properties
 of the Whittaker's functions $M(\tilde{a},\tilde{b},\zeta)
 \sim\zeta^{\tilde{a}-\tilde{b}}e^{\zeta}\Gamma(\tilde{b})/\Gamma(\tilde{a})$ and $U(\tilde{a},
 \tilde{b},\zeta)\sim \zeta^{-\tilde{a}}$,
 one can get the large $r$ behavior of the far-region solution as
 \begin{eqnarray}
 R\sim \alpha\frac{\Gamma(2l+2)}{\Gamma(l+1-\rho)}(2j\omega r^2)^{-1-\rho}e^{j\omega r^2}
 +\beta(2j\omega r^2)^{-1-\rho}e^{-j\omega r^2}\;.
 \end{eqnarray}
 Obviously the first term is divergent at asymptotic infinity. To match the Dirichlet
 boundary condition at infinity, we have to set $\alpha=0$. Thus the far-region solution
 with the Dirichlet boundary condition at asymptotic infinity is given by
 \begin{eqnarray}
 R=\beta(2j\omega)^l r^{2l} e^{-j\omega r^2} U(l+1-\rho,2l+2,2j\omega r^2)\;.
 \end{eqnarray}
 This solution is just the solution of scalar field wave equation in the background of the pure five-dimensional G\"{o}del spacetime \cite{konoplyaads,Hiscock,tachyon}.

 We assume for a moment that we have no black hole, and calculate the real
 frequencies that can propagate in the pure five-dimensional G\"{o}del spacetime.
 In this setup, the spacetime geometry is horizon-free, and the solution of
 the scalar field perturbation in the background of the pure five-dimensional G\"{o}del spacetime
 should be regular at the origin $r=0$. When $\zeta\rightarrow 0$,
 using the properties of Whittaker's function
 $U(\tilde{a},\tilde{b},\zeta)\sim\zeta^{1-\tilde{b}}\Gamma(\tilde{b}-1)/\Gamma(\tilde{a})$,
 one can get the small $r$ behavior of the far-region solution as
 \begin{eqnarray}
 R\sim \beta(2j\omega)^{-l-1}\frac{\Gamma(2l+1)}{\Gamma(l+1-\rho)}r^{-2l-2}\;.
 \end{eqnarray}
 So, when $r\rightarrow 0$, $r^{-2l-2}\rightarrow\infty$, and the solution diverges.
 To have a regular solution at the origin $r=0$, we must demand that $\Gamma(l+1-\rho)
 \rightarrow\infty$. This occurs when the argument of the gamma function is a non-positive
 integer. Therefore, we have the condition
 \begin{eqnarray}
 l+1-\rho=-N,\;\;\;N=0,1,2,\cdots\;.
 \end{eqnarray}
 So the requirement of the regularity of the wave solution at the origin
 selects the frequencies of the scalar field that might propagate in the
 pure five-dimensional G\"{o}del spacetime
 \begin{eqnarray}\label{normalmode}
 \omega_N=8j(N+l+m+1)\;.
 \end{eqnarray}

 Now let us come back to the Kerr-G\"{o}del black hole case. In the spirit of
 \cite{cardoso2004ads}, we expect that there will be a small imaginary part $\delta$ in the allowed
 frequencies induced by the black hole event horizon
 \begin{eqnarray}
  \omega=\omega_N+i\delta\;.
 \end{eqnarray}
 From $\Psi\sim e^{-i\omega t} $, one can see that the small imaginary
 $\delta$ describes the slow growing instability
 of the modes when $\delta>0$. Our task is to prove that
 $\delta$ is positive in the regime of superradiance.

 Inserting this expression for the frequency $\omega$, one can get the
 far-region solution of the radial wave equation as
 \begin{eqnarray}\label{farsolution}
 R=\beta(2j\omega)^l r^{2l} e^{-j\omega r^2} U(-N-i\delta/8j,2l+2,2j\omega r^2)\;.
 \end{eqnarray}
 In order to match the far-region solution with the near-region solution,
 we need to find the small $r$ behavior of the far-region solution.
 It is known that the Whittaker's function $U(\tilde{a},\tilde{b},\zeta)$ can be
 expressed in terms of the Whittaker's function $M(\tilde{a},\tilde{b},\zeta)$.
 By inserting this relation on the far-region solution (\ref{farsolution}),
 we can show that the far-region solution can be rewritten as
 \begin{eqnarray}
 R=\beta(2j\omega)^l r^{2l} e^{-j\omega r^2}
 \frac{\pi}{\sin\pi(2l+2)}\left[
 \frac{M(-N-i\delta/8j,2l+2,2j\omega r^2)}
 {\Gamma(-N-2l-1-i\delta/8j)\Gamma(2l+2)}
 \right.\nonumber\\
 \left.-(2j\omega)^{-2l-1}r^{-4l-2}
 \frac{M(-N-2l-1-i\delta/8j,-2l,2j\omega r^2)}
 {\Gamma(-N-i\delta/8j)\Gamma(-2l)}
 \right] \;.
 \end{eqnarray}

 Applying to this expression the functional expressions
 for the gamma functions
 \begin{eqnarray}
 \Gamma(n+1)&=&n!\;,\nonumber\\
 \Gamma(z)\Gamma(1-z)&=&\frac{\pi}{\sin\pi z}\;,
 \end{eqnarray}
 it is easy to show that
 \begin{eqnarray}
 \frac{1}{\Gamma(-2l)}=-\frac{\sin\pi(2l+2)}{\pi}(2l)!\;,
 \end{eqnarray}
 and for the small $\delta$
 \begin{eqnarray}
 \frac{1}{\Gamma(-N-2l-1-i\delta/8j)}&=&(-1)^N \frac{\sin\pi(2l+2)}{\pi}
 (N+2l+1)!\;,\nonumber\\
 \frac{1}{\Gamma(-N-i\delta/8j)}&=&(-1)^{N+1}N!i\delta/8j\;.
 \end{eqnarray}

 Then by using the property of Whittaker's function $M(\tilde{a},\tilde{b},0)=1$,
 one can get the small $r$ behavior of the far-region solution as
 \begin{eqnarray}\label{farsolutionsmall}
 R=\beta(-1)^N(2j\omega_N)^l \left[
 \frac{(N+2l+1)!}{(2l+1)!}r^{2l}
 -i\delta \frac{(2l)!N!}{2^{2l+4}j^{2l+2}\omega_N^{2l+1}}r^{-2l-2}
  \right] \;.
 \end{eqnarray}

 \subsection{Matching condition: the unstable modes}

 By comparing the large $r$ behavior of the near-region solution with
 the small $r$ behavior of the far-region solution, one can conclude
 that there exists the overlapping region $M\ll x-x_+\ll1/\omega$
 where the two solutions should match. In this region,
 the matching of the near-region solution in the large $r$ region
 (\ref{nearsolutionlarge}) and
 the far-region solution in the small $r$ region
 (\ref{farsolutionsmall}) yields
 the allowed values of the small imaginary part $\delta$
 in the frequency $\omega$
  \begin{eqnarray}
  \delta\cong -\sigma(\omega_N-m\Omega_H)
  r_+(r_+^4+2Ma^2)^{1/2}(r_+-r_-)^{2l}j^{2l+2}\;,
  \end{eqnarray}
 where
  \begin{eqnarray}
  \sigma&=&2^{2l+4}\omega_N^{2l+1}
  \frac{(l!)^2(2l+1+N)!}{((2l)!(2l+1)!)^2N!}
  \left[\prod_{k=1}^{l}(k^2+4\varpi^2)\right]\;,
  \end{eqnarray}
 with $\varpi=(\omega_N-m\Omega_H)r_+(r_+^4+2Ma^2)^{1/2}/2(r_+^2-r_-^2)$.
 So, we have
  \begin{eqnarray}
   \delta\propto -(Re[\omega]-m\Omega_H)\;.
  \end{eqnarray}
 It is easy to see that, in the superradiance regime, $Re[\omega]-m\Omega_H<0$,
 the imaginary part of the complex frequency
 $\delta>0$. The scalar field has the time dependence
 $e^{-i\omega t}=e^{-i\omega_N t}e^{\delta t}$, which implies
 the exponential amplification of superradiance modes.
 This will lead to the instability of these modes.

 From the normal modes in pure five-dimensional G\"{o}del
 spacetime (\ref{normalmode}), we can see that $Re[\omega]\sim j$.
 We have assumed that $\omega M\ll 1$. So we have $jM\ll 1$, which
 is consistent with the small G\"{o}del black hole assumption
 $jr_+\ll 1$ because $M\sim r_+^2$. This is to say that the two assumptions
 we have made in this section are consistent with each other.

 Let us make a qualitative comparison of our analytical results with the
 numerical one in \cite{knopolya}. From Eq.(46), we can see that the
 growth rate $\delta$ of the superradiant modes are propotional to $j^{2l+2}$.
 This implies that the larger $j$ corresponds to the higher superradiant
 instability growth rate, which supports the numerical conclusion in \cite{knopolya}.
 Next, the superradiant condition that $Re[\omega]-m\Omega_H<0$ can
 make a further limit on the parameter space where the superradiance can occur.
 From Eq.(38), we can see the superradiance condition
 for the mode of $l=m=1$ and $N=0$ becomes $24j<\Omega_H$. Because we are working on the parameter space that $jr_+^2\ll 1$, we can further take the limit $jM\ll 1$ in the expression of event horizon Eq.(4) and angular momentum Eq.(8). So the limit on the parameter $j$ can be approximated by the following expression
 \begin{eqnarray}
 j\sqrt{M}<\frac{1-\sqrt{1-2(a/\sqrt{M})^2}}{24(a/\sqrt{M})}\;,
 \end{eqnarray}
 where, the parameter $a/\sqrt{M}$ takes the value in the region of $(0,0.71)$, otherwise
 there will be no black hole in spacetime. By submitting $a/\sqrt{M}=0.71$ into the inequality,
 one can see that for $j\sqrt{M}\sim 0.059$ there is no superradiance. The numerical one in \cite{knopolya} is $j\sqrt{M}\sim 0.075$. The region of the superradiance in the parameter space
 by using the analytical method roughly coincides with the numerical results in \cite{knopolya}.
 So, comparing our result with the numerical one in \cite{knopolya}, it can be seen that our results
 are rough and can reproduce the numerical conclusion partly. The analytical method can not solved
 the superradiance instability as precisely as the numerical method.

 At last, we can conclude that the five-dimensional small
 Kerr-G\"{o}del black hole is unstable against the massless
 scalar field perturbation. This instability is caused by
 the superradiance of the scalar field.

  \section{Conclusion}

 This paper is devoted to an analytical study of superradiant
 instability of five-dimensional small Kerr-G\"{o}del black hole.
 This instability has been found by R. A. Konoplya and A. Zhidenko
 using the numerical methods previously in \cite{knopolya}.
 Generally, superradiant instability naturally happens
 when two conditions are satisfied: (1) Black hole has rotation;
 (2) There is a natural reflecting mirror outside the black hole.
 In the present case, Dirichlet boundary condition
 at infinity for the asymptotically G\"{o}del black hole, which is obtained
 in section 3 by analogy with the AdS background,
 plays the role of reflecting mirror.
 We have adopted the analytical methods which is used in \cite{cardoso2004ads}
 to study the superradiant instability of small Kerr-AdS black hole.
 We assume that the energy of scalar field perturbation is low
 and the scale of G\"{o}del black hole is small.
 Then we divide the space outside the event horizon into
 the near-region and the far-region and employ the matched asymptotic
 expansion method to solve the scalar field wave equation.
 The analysis of complex quasinormal modes explicitly
 shows that the complex parts are positive in the regime of superradiance,
 which implies the growing instability of these modes.
 This is to say that the five-dimensional small Kerr-G\"{o}del black hole
 is unstable against scalar field perturbation.

 As is well-known, the gravitational perturbation will also undergo
 an superradiant amplification when scattered by the event horizon.
 Unlike the simply rotating Kerr-AdS black hole, where the superradiant
 instability of a massless scalar field implies the gravitational instability
 \cite{kodama}, we can not obtain the similar
 conclusion for asymptotically G\"{o}del black hole
 directly. So it will be interesting to study the gravitational (in)stability of
 five-dimensional Kerr-G\"{o}del black hole. Because of the complexity
 of the metric and the field equation, it will be a challenging project for
 future work.

 \section*{Acknowledgement}

 The author would like to thank Ming-Fan Li for reading the
 manuscript and useful comments. This work was supported by
 NSFC, China (Grant No. 11147145 and No. 11205048).


\begin{thebibliography}{99}

 \bibitem{zeldovich}

 Ya. B. Zel'dovich, Pis'ma Zh. Eksp. Teor. Fiz. \textbf{14}, 270
 (1971) [JETP Lett. \textbf{14}, 180 (1971)]; Zh. Eksp. Teor. Fiz.
 \textbf{62}, 2076 (1972) [Sov. Phys. JETP \textbf{35}, 1085 (1972)].

 \bibitem{bardeen}

 J. M. Bardeen, W. H. Press, and S. A. Teukolsky,
 Astrophys. J. \textbf{178}, 347 (1972).

 \bibitem{misner}

 C. W. Misner, Bull. Am. Phys. Soc. \textbf{17}, 472(1972).

 \bibitem{starobinsky}

 A. A. Starobinsky, Zh. Eksp. Teor. Fiz. \textbf{64}, 48 (1973) [Sov.
 Phys. JETP \textbf{37}, 28 ( 1973)]; A. A. Starobinsky and S. M.
 Churilov, Zh. Eksp. Teor. Fiz. \textbf{65}, 3 (1973) [Sov. Phys.
 JETP \textbf{38}, 1 (1973)].


 \bibitem{press}

 W. H. Press, and S. A. Teukolsky, Nature (London) \textbf{238},
 211 (1972).

 \bibitem{cardoso2004bomb}

 V. Cardoso, O. J. C. Dias, J. P. S. Lemos, and S. Yoshida,
 Phys. Rev. D \textbf{70}, 044039 (2004).

 \bibitem{Hod}

 S. Hod, and O. Hod, Phys. Rev. D \textbf{81}, 061502(2010).

 \bibitem{Rosa}

 J. G. Rosa, JHEP \textbf{1006}, 015(2010).

 \bibitem{Lee}

 J.-P. Lee, Mod. Phys. Lett. A \textbf{27}, 1250038(2012).


 \bibitem{kerrunstable}

 T. Damour, N. Deruelle, and R. Ruffini, Lett. Nuovo Cimento
 Soc. Ital. Fis. \textbf{15}, 257(1976).

 \bibitem{detweiler}
 S. Detweiler, Phys. Rev. D \textbf{22}, 2323(1980); T. M. Zouros, and
 D. M. Eardley, Ann. Phys. (N.Y.) \textbf{118}, 139(1979); H. Furuhashi
 and Y. Nambu, Prog. Theor. Phys. \textbf{112}, 983(2004).


 \bibitem{dolan}

 M. J. Strafuss and G. Khanna, Phys. Rev. D \textbf{71}, 024034(2005);
 S. R. Dolan, Phys. Rev. D \textbf{76}, 084001(2007).

  \bibitem{hodPLB2012}

 S. Hod, Phys. Lett. B \textbf{708}, 320(2012).

 \bibitem{konoplyaPLB}

 R. A. Konoplya, Phys. Lett. B, \textbf{666}, 283(2008).

 \bibitem{DiasPRD2006}

 O. J. C. Dias, Phys. Rev. D \textbf{73}, 124035(2006).

 \bibitem{cardoso2004ads}

 V. Cardoso, and O. J. C. Dias, Phys. Rev. D \textbf{70}, 084011(2004).

 \bibitem{cardoso2006prd}

 V. Cardoso, O. J. C. Dias, and S. Yoshida, Phys. Rev. D \textbf{74}, 044008(2006).


 \bibitem{aliev}

 A. N. Aliev, and O. Delice, Phys. Rev. D \textbf{79}, 024013(2009).

 \bibitem{uchikata}

 N. Uchikata, and S. Yoshida, Phys. Rev. D \textbf{83}, 064020(2011).

 \bibitem{ranliplb}

 R. Li, Phys. Lett. B \textbf{714}, 337(2012).

 \bibitem{clement}

 G. Clement, D. Galtsov and C. Leygnac,
 Phys. Rev. D \textbf{67}, 024012(2003).

 \bibitem{knopolya}

 R. A. Konoplya, and A. Zhidenko, Phys. Rev. D \textbf{84}, 104022(2011).


 \bibitem{gimon}

 E. G. Gimon and A. Hashimoto, Phy. Rev. Lett. \textbf{91}, 021601(2003).

 \bibitem{godel}

 K. G\"{o}del, Rev. Mod. Phys. \textbf{21}, 447(1949).


 \bibitem{banichconpere}

 G. Barnich and G. Compere, Phys. Rev. Lett. \textbf{95}, 031302(2005).

 \bibitem{wupengprd}

 S. Q. Wu and J. J. Peng, Phys. Rev. D \textbf{83}, 044028(2011).

 \bibitem{scalargodel}

 R. A. Konoplya and E. Abdalla, Phys. Rev. D \textbf{71}, 084015(2005);
 S. B. Chen, B. Wang, and J. L. Jing, Phys. Rev. D \textbf{78},064030(2008);
 X. He, B. Wang, and S. B. Chen, Phys. Rev. D \textbf{79}, 084005(2009);
 W. Li, L. Xu, and M. Liu, Class. Quant. Grav. \textbf{26}, 055008(2009).

 \bibitem{wu}

 S. Q. Wu, Phys. Rev. Lett. \textbf{100}, 121301(2008).

 \bibitem{konoplyaads}

 R. A. Konoplya, and A. Zhidenko, Phys. Rev. D \textbf{84}, 064028(2011).

 \bibitem{Hiscock}

 W. A. Hiscock, Phys. Rev. D \textbf{17}, 1497(1978);
 S. N. Guha Thakurta, Phys. Rev. D \textbf{21}, 864(1980).

 \bibitem{tachyon}

 R. A. Konoplya, Phys. Lett. B \textbf{706}, 451(2012);
 R. A. Konoplya, and A. Zhidenko, arxiv: 1110.2015[hep-th].


 \bibitem{kodama}

 H. Kodama, Prog. Theor. Phys. Suppl. \textbf{172}, 11(2008); arxiv:0711.4184[hep-th].


\end{thebibliography}
 \end{document}